\documentclass[12pt]{iopart}
\usepackage{graphics,graphicx,epsfig,latexsym,amssymb,verbatim,color,bm}
\newcommand{\be}{\begin{eqnarray}}
\newcommand{\ee}{\end{eqnarray}}

\newcommand{\ket}[1]{\vert #1\rangle}

\usepackage{theorem}

\def\squareforqed{\hbox{\rlap{$\sqcap$}$\sqcup$}}
\def\qed{\ifmmode\squareforqed\else{\unskip\nobreak\hfil
\penalty50\hskip1em\null\nobreak\hfil\squareforqed
\parfillskip=0pt\finalhyphendemerits=0\endgraf}\fi}
\def\endenv{\ifmmode\;\else{\unskip\nobreak\hfil
\penalty50\hskip1em\null\nobreak\hfil\;
\parfillskip=0pt\finalhyphendemerits=0\endgraf}\fi}

\begin{document}
\title{Noise resistance of violation of local causality for pure three-qutrit entangled states}
\author{Wies{\l}aw Laskowski, Junghee Ryu, Marek \.Zukowski}
\address{Institute of Theoretical Physics and Astrophysics, University of Gda\'nsk, 80-952 Gda\'nsk, Poland}

\begin{abstract}
Bell's Theorem started with two qubits (spins $1/2$). It is a no-go statement on classical (local causal) models of quantum correlations. Only after 25 years, it turned out that for three qubits the situation  is even more mind boggling. General statements concerning higher dimensional systems, qutrits etc.,  started to appear even later, once the spin (higher than $1/2$) picture was replaced by a more broad one, allowing all possible observables. This work is continuation of the Gdansk effort to take advantage of the fact that Bell Theorem can be put in form of a linear programming problem, which in turn can be translated into a computer code. Our results are numerical and classify the strength of violation of local causality by various families of three-qutrit states, as measured by resistance to noise. This is a  previously uncharted territory. The results may be helpful in suggesting which three-qutrit states are handy for applications in quantum information protocols.  One of the surprises is  that the W state turns out to reveal a stronger violation of local causality than the GHZ state.
\end{abstract}
\pacs{03.65.Ud, 03.67.Mn}

\section{Introduction}
Bell's Theorem ~\cite{Bell64}, celebrated in this Volume, is one of the great steps in our understanding of the physics of the micro-world. It signals a much greater departure from the classical picture than the uncertainty relations (note that e.g., frequency-time uncertainty holds for classical waves). The non-commutative nature of quantum mechanics and the collapse postulate clearly depart from the classical description, but where are phenomena which directly show the impossibility of a return to a picture suggested by classical mechanics and field theory?  Well, this is the Bell Theorem for correlations of entangled systems. The theorem tells us that in such a case we cannot have a local causal model, obeying all rules of classical (Kolmogorov) probability theory.\footnote{There are no other tools in quantum mechanics than quantum states, observables, and the principle of unitary evolution and Born probabilistic rule using all that. Thus, for example, if a single photon impinges on a $50/50$ beam splitter there is no direct cause in the formalism telling whether it will be reflected or transmitted. Adding to the description any new parameters outside of quantum mechanics as causes of this or that behavior is in effect introduction of hidden variables (hidden, as they are not present in quantum formalism).}   
Bell's Theorem is a mathematical property of quantum mechanical description of Nature. However, the universal form of Bell inequalities, in the modern parlance a device independent one, allows one to ask whether Nature itself is local causal. Experiments point into a negative answer, but the matter is still unsettled (for a review of photonic experiments see \cite{PAN}).  Once a single loophole free experiment is done the fundamental question would be answered. Still Bell inequalities have a utilitarian aspect. They point to strictly nonclassical phenomena, which as they are impossible to realize with classical means, can lead to new practical applications. Thus apart from fundamental importance, continuation of the studies related to Bell Theorem is a search for new nonclassical phenomena, states, and quantum tools for future technologies.

This work will study an important characteristics of entangled states - noise resistance of violation of local causality. The quantum correlations are fragile, can be totally blotted out by noise, which is inevitable in any experiment, and thus in all applications. Therefore,  one should have estimates of some critical noise parameters to see usefulness of quantum correlations characteristic for a given state.  The states which we shall analyze here go beyond the standard Bell Theorem. They are three particles (subsystem) ones, and they are for systems more complicated than qubits. Such states allow new applications, impossible with two-qubit entanglement of the pioneering works (see e.g. \cite{DURT}).

Surprisingly, the first paper showing the extraordinary features of three or more qubits, in a maximally entangled state, appeared 25 years after Bell's work, \cite{GHZ}. For more than a generation nobody cared to investigate this new realm (with an exception of \cite{Svetlichny87}, however the striking properties of perfect correlations were left undiscovered). Three or more particle correlations lead to new possibilities in applications, like three-party cryptography, with the third party holding a key to the key of the other two (``secret sharing") \cite{ZZWH}, or reduction of communication complexity of some computational tasks \cite{BUHRMAN, BZZP}.

Studies of higher dimensional systems had also a slow start. This was due to the fact that earlier investigations treated such systems as spins  higher than $1/2$, and employed only observables which are components of the spin operators. Thus, not all unitary transformations between measurement bases were allowed. What is perhaps even more important, the spin components are observables with classical counterparts. Introduction to the discussion of new observables,  in form of multiport interferometers \cite{ZEILINGER}, and subsequent generalizations \cite{RECK, TRITTER}, allowed to show that the violation of the Bell inequalities is stronger for higher dimensional systems than for qubits \cite{KASZLIKOWSKI} (this work was numerical, analytic reproduction of the results was given in \cite{KASZ, Collins02}). This discovery was a signature of possible practical advantages related with higher dimensional systems (e.g., higher security in quantum cryptography~\cite{CERF}).

Noise resistance of violation of local causality by quantum correlations was introduced in Ref. \cite{KASZLIKOWSKI} to quantify the strength of violation of local causality by quantum correlations.  Previously in various works authors were using such parameters as violation factors or numerical differences between a Bell bound and the quantum value. However such measures do not allow comparison between different Bell inequalities, or versions of Bell Theorem which are not based directly on them, but rather on numerical methods based on linear programming, like in \cite{KASZLIKOWSKI}. As an illustrative example notice that any violation of the Clauser-Horne \cite{CH} inequality is always by an infinite factor (as its local causal bound is zero).

Noise resistance depends on the model used. The simplest one is a ``white noise" admixture, see later. Generally one can define a ``critical visibility" \cite{KASZLIKOWSKI}, as the value of the parameter $v$, nonnegative, less or equal to one, for which, under a fixed set of conditions (like number of settings, types of observables) a mixed state 
defined by 
\begin{equation} 
\varrho=v\varrho_{state} + (1-v) \varrho_{noise}
\end{equation}
loses uniquely quantum properties of the original state $v\varrho_{state}$
 (the symbol $ \varrho_{noise}$ stands for the noise admixed to the state, it is a form of a certain separable density operator, the actual form of which depends on the chosen noise model).

In this work we attempt to obtain noise resistance values for three qutrit entangled states. This is done using numerical methods, as in \cite{KASZLIKOWSKI}. Only some families are analyzed. This is due to the fact that, despite our efforts (for a computer science outline of the current upgraded numerical method see \cite{SR3}), the machine time to perform the numerical calculations is quite long.  

Let us briefly sketch related earlier developments.
Acin {\em et al.} constructed a tight Bell inequality of a three-qutrit system~\cite{Acin04}. They showed that the maximal violation is for maximally entangled GHZ state, with two measurements for each observer defined by three-port beam splitters (tritters)~\cite{TRITTER}. Moreover, the inequality reproduced a numerical result of the resistance to white noise, with critical visibility $v_{crit}=0.6$, which was obtained by numerical methods in ~\cite{KASZLIKOWSKI43}.
In Ref.~\cite{Deng09}, Deng {\em et al.} studied the quantum violations of three-qutrit GHZ states by using their Bell-type inequality, based on Svetlichny's ideas ~\cite{Svetlichny87}. In case of maximally entangled GHZ states, no violation was reported. 
In Ref.~\cite{Lim10} a set of Bell inequalities was derived  for three-qutrit systems involving three measurement settings  for each subsystem. We have checked that the critical visibility for the inequalities is about $0.75$ (all value quotations are for white noise).
Grandjean {\em et al.} suggested a family of Bell inequalities for tripartite and high-dimensional systems with two measurement settings~\cite{Grandjean12}. For the two-dimensional case, they can be reduced to the Mermin-Bell~\cite{Mermin90PRL} inequality, and merge with
 the CGLMP inequality~\cite{Collins02} for the bipartite case. Numerical results in terms of the critical visibility to white noise were shown for several tripartite states, such as a maximally entangled GHZ state, W state, and the three-qutrit singlet state (also known as  Aharonov state). The critical visibility for the GHZ state was $0.75$, while  for the three-qutrit singlet state $0.7846$~\cite{Grandjean12}.

\section{Description of the method}

In our numerical analysis (called {\scshape steam roller}) we consider
some classes of pure states of three qutrits.
Three spatially separated observers perform measurements of
$m$ alternative local three valued observables:
$A_1, A_2, \dots ,A_m$ for Alice, $B_1, B_2, \dots , B_m$ for Bob, and $C_1, C_2, \dots , C_m$ for Charlie ($m=2,3$). We
assume that they measure observables which belong to the following families: 

(i) U(3) measurements, by this we mean the full family of observables, as defined by orthogonal projectors (the symbol U(3) signifies the fact that all such observables are linked by U(3) type unitary transformations),

 (ii) T -- denotes a subset of the above observables which are defined by unitary transformations, starting form the computational basis, which use one unbiased three-port beam splitter (tritter, see \cite{TRITTER}) performing a unitary transformation given by  the matrix $U_{ij} = e^{2 i j \pi/3}/\sqrt{3}$, and a set of phase shifters in front of it.  This class was singled out, as it played an important role in breaking the limitations linked with spin observables, discussed in Introduction. Also, such observables are rapidly becoming more feasible using the modern methods of integrated optics \cite{SCH}.

By saying that an experiment is local causal  (realistic) we understand that it has a local realistic model, 
which is equivalent to the existence of a joint probability distribution $p_{lr}(a_1,...,a_m,b_1,...,b_m,c_1,...,c_m)$,
where $a_i (b_i,c_i) =\{0,1,2\}$ denotes the results of the measurement of Alice's (Bob's, Charlie's) $i$th observable. Quantum predictions for the probabilities should be given, if the model exists, by the marginal sums:
\begin{equation}
P(a_i,b_j,c_k|A_i,B_j,C_k) = \sum_{a_{i'},b_{j'},c_{k'}=0} p_{lr}(a_1,..., a_m,b_1,...,b_m,c_1,...,c_m), 
\label{constraints}
\end{equation}
where $P(a_i,b_j,c_k|A_i,B_j,C_k)$ denotes the probability of obtaining the result $a_i$ by Alice, $b_j$ by Bob and $c_k$ by Charlie if they measure  observables $A_i$, $B_j$ and $C_k$, respectively, and $i' \neq i, j' \neq j, k' \neq k$.

If we admix to the three qutrit state the ``white" noise\footnote{in our analysis we also consider a product noise admixture described in Sec. \ref{sec:product}}, in the form of  a completely random process, the quantum probabilities $P_v(a_i,b_j,c_k|A_i,B_j,C_k)$ are given by
\begin{equation}
P_v(a_i,b_j,c_k|A_i,B_j,C_k) = v P(a_i,b_j,c_k|A_i,B_j,C_k) + \frac{1-v}{27},
\end{equation}
where $P(a_i,b_j,c_k|A_i,B_j,C_k)$ is the quantum prediction for the state without the noise admixture. For such noisy states there are always exists the critical parameter $v_{crit}$, such that for $v<v_{crit}$ there exist a local realistic joint probability distribution $p_{lr}$ which satisfies the set of constraints (\ref{constraints}). Our task, for a given $\rho$, is to find the critical parameter $v_{crit}$. This can be done by means of linear programming. 

The numerical procedure was first used in \cite{BATURO}, and further developed in Ref. \cite{KASZLIKOWSKI} to show that violation of local realism is stronger (more white noise resistant) for two qu$N$its than for two qubits and that this increases with $N$. Later the method was also used for analysis of violations of local realism by two qutrits in all possible pure
entangled states \cite{SR2} and  different types of qubit quantum states, which are often discussed in the context of quantum information \cite{SR1}. The method is described in details in Refs. \cite{SR1, SR3}.

\section{Results}

We applied the numerical method to different types of
three qutrit quantum states, namely, the generalized GHZ state, a family of the Dicke states and the singlet state. The critical visibility for two and three settings per side was determined. Selected results are presented below \footnote{The optimal measurement settings for the most important examples are shown in the Appendix}.

\subsection{Generalized GHZ states}

We considered the following a class of pure states of three qutrits (generalized GHZ states):
\begin{equation}
\ket{\psi(\alpha)} = \cos\alpha \ket{000} + \frac{1}{\sqrt{2}}\sin \alpha (\ket{111} + \ket{222}).
\label{ghzstate}
\end{equation}
We have found a range of the $\alpha$ parameter of ($\ref{ghzstate})$, for which the critical visibility is lower than the lowest previously  known critical visibility for white noise (0.43) for three qutrit states, given in Ref. \cite{KASZLIKOWSKI43}. The lowest critical visibility resulting from our analysis is now  0.4256, and  corresponds to the rank-2 state $\ket{\psi(90^o)}$. A similar effect was observed for two qutrits and reported in Refs. \cite{SR2, GISIN}. This phenomenon is visible if one uses the most general local three-dimensional observables [U(3)], whereas for the observables based on  transformations T it is unobservable. 

The lowest critical visibility for transformations T only and two measurement settings per side (0.5931) is achieved for an asymmetric state $\ket{\psi_{asym}}=\ket{\psi(\sim{50^o})}$. For the symmetric state $\ket{\psi_{sym}}=\ket{\psi(54.74^o)}=\frac{1}{\sqrt{3}}(\ket{000}+\ket{111}+\ket{222})$ the critical visibility is equal to 0.6000 as in Refs. \cite{Acin04, KASZLIKOWSKI43}. Surprisingly for three measurement settings per side the plot of the critical visibility in a function of $\alpha$ becomes more symmetric. The minimal critical visibility (0.5000) is achieved by the symmetric proper three-qutrit
GHZ  state $\ket{\psi_{sym}}$. This is exactly the same as the well known value for the case of a three-qubit GHZ state, and two settings per party. 

There are some ranges of the parameter $\alpha$, for which the observables
generated by U(3) give a better critical visibility for three measurement settings per side than the one obtained for two measurement settings ($46^o \leq \alpha \leq 58^o$). An example is the state $\ket{\psi_{sym}}$, for which the critical visibilities for two and three measurements settings are equal to 0.5264 and 0.5000, respectively.

In the case of the two-qutrit GHZ state, after increasing the number of settings to three one cannot observe reduction of the critical visibility \cite{SR2}.

\begin{figure}[!h]\center
\resizebox{9cm}{!}{
\includegraphics{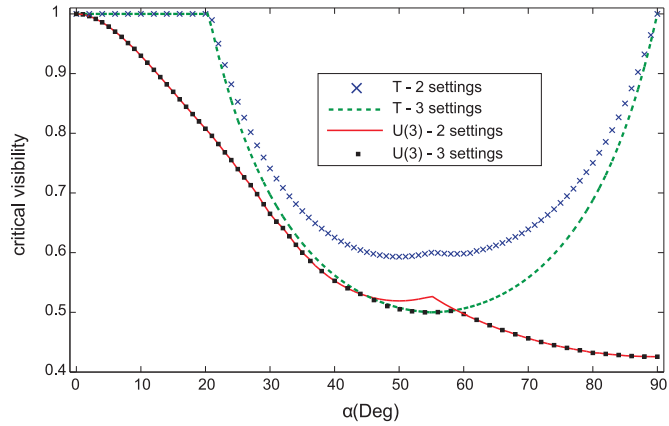}}
\caption{\label{} The critical visibilities for the states $\ket{\psi(\alpha)}$, and white noise. Each line
corresponds to a different type of operator family (U(3), T) and the number of  measurement settings per observer ($m=2,3$).
The ``discontinuities" of the ``derivatives" of the curves are interpreted as points at which another Bell inequality takes over as the one which is  violated with highest noise resistance (``most strongly").}
\label{Fig:3qutrits}
\end{figure}

\label{sec:product}

\begin{figure}[!h]\center
\resizebox{9cm}{!}{
\includegraphics{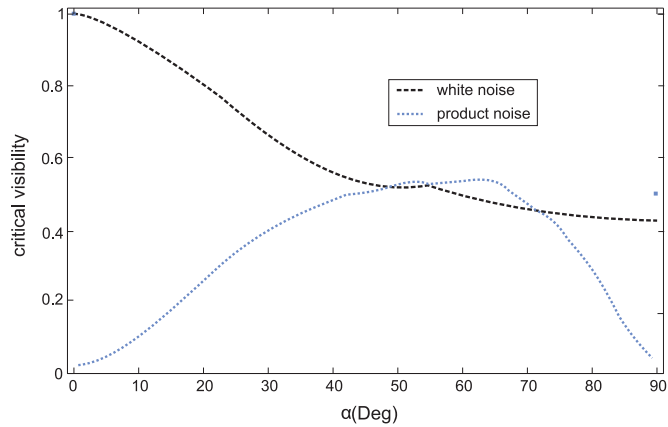}}
\caption{
The critical visibilities $v_{crit}$ for the states $\ket{\psi(\alpha)}$ for the case in which all observers
use any observables (U(3) case) and two measurement settings. The dotted line (blue) corresponds to the product noise admixture and it is compared with the results for white noise (dashed, black). Note the "singular" point for $90^o$ with visibility $1/2$, for the product noise. This is for the rank-2 state (effectively, a three qubit GHZ state). Also note that in the case of a perfect three-qutrit GHZ state product noise and white noise are described by identical operators. This is reflected in the figure by the mid crossing point of the product and white noise curves. The other crossing points seem accidental.}
\label{Fig:prod}
\end{figure}

We have  also studied  a {\em product} noise admixture
of the form $v \rho + (1-v) \rho_A \otimes \rho_B \otimes \rho_C$, where $\rho_{i}$ is the reduced density matrix
of  system $i$. The results, compared with the  white noise ones, are shown
in Fig. \ref{Fig:prod}. 
Note that the state $\ket{\psi(0)}$ (product state) does not violate local realism, whereas for $\alpha$ close to 0 the state (\ref{ghzstate}) violates local realism for visibility approaching $0$. A  situation of a different kind is for $\alpha=90^{o}$. The critical visibility for the rank-2 state $\ket{\psi(90^{o})}$ is equal 0.5 and immediately drops to almost $0$ in the neighborhood of $90^{o}$. 

\subsection{Dicke states}

We also tested a family of the three qutrit Dicke states:
\begin{eqnarray}
\ket{D_{3}^{1}}&=& \frac{1}{\sqrt{3}}(\ket{001} + \ket{010} + \ket{100}), \\
\ket{D_{3}^{2}}&=& \frac{1}{\sqrt{15}} \left(\ket{002} + \ket{020} + \ket{200} + 2\left(\ket{011} + \ket{101} + \ket{110}\right)\right),\\
\ket{D_{3}^{3}}&=& \frac{1}{\sqrt{10}}(\ket{012} + \ket{021} + \ket{102} + \ket{120} + \ket{201} + \ket{210} + 2 \ket{111}).
\end{eqnarray}
The results for U(3) measurements are presented in Table \ref{tab:dicke}.  The most robust quantum properties are observed for the state $\ket{D_{3}^{1}}$, which is simply the qubit-like W state. The state is more robust against white noise than the symmetric GHZ state. Moreover, the state violates local realism for any admixture of the product noise. Also, the qutrit Dicke states are more noise resistant than  three qubit W state, for which the two setting value is  0.6442, while the three setting one is 0.6330. But the most surprising result is that the quantum nature of correlations of the state $|D_{3}^{1}\rangle$ is totally resistant to the product noise (critical visibility zero), whereas for a three qubit W state has 0.662 (two settings per observer ) 0.660 (three settings).

\begin{table}
\caption{\label{tab:dicke} Critical visibilities for the three qutrit Dicke states, two and three measurement settings, white and product noise admixtures.}
\footnotesize\rm
\begin{tabular*}{\textwidth}{@{}l*{15}{@{\extracolsep{0pt plus12pt}}l}}
\br
State & \multicolumn{2}{c}{white noise}& \multicolumn{2}{c}{product noise}\\
      & 2 settings & 3 settings & 2 settings & 3 settings \\
\mr
$|D_{3}^{1}\rangle$ & 0.4993 & 0.4866& 0.0& 0.0\\
$|D_{3}^{2}\rangle$ & 0.5363 & 0.5285& 0.5925& 0.5894\\
$|D_{3}^{3}\rangle$ & 0.5338 & 0.5108& 0.5409& 0.5409\\
\br
\end{tabular*}
\end{table}

\subsection{Singlet state}

Finally, we analyzed the three qutrit Aharonov singlet state \cite{singlet}:
\begin{eqnarray}
\ket{\rm{singlet}}=\frac{1}{\sqrt{6}}\left(\ket{012} - \ket{021} -\ket{102} +\ket{120} +\ket{201} - \ket{210}\right).
\end{eqnarray}
Surprisingly, the obtained critical visibilities for all U(3) type observables for both white and product noise are the same and read 0.6456, for two measurement settings, and 0.6219, for three settings.

\section{Conclusions}
The numerical method presented here once more proved to be a useful tool for seeking strictly quantum correlations in situations in which the full set of tight Bell inequalities is either unknown, or is too vast to be useful. The obtained values of critical visibility are either the best so far, or apply to previously unstudied cases.  
Note that the numerical code which we used is equivalent to the full set of tight Bell inequalities for the given problem. The Bell inequalities that are violated by the states are not specified. They can be obtained by using the method of Ref. \cite{Elliott}.

We observe that in many cases allowing three settings implies a lower threshold visibility. Some results are very surprising, like the full noise resistance of the Dicke W state, in the case of product noise. Also this state beats the three-qutrit  GHZ state in resistance with respect to white noise. However this should not be upsetting news for the authors of the GHZ paper, as they were the first ones to discuss W states \cite{ZHG} in the context of quantum correlations. Note that an earlier result showed that three qubit W states beat the original GHZ ones in white noise resistance, if the number of subsystems is greater than ten \cite{ADITI}.

All these results suggest that in the case of three qutrit states it might be the case that possible applications in quantum information protocols could be most exciting in the case of the W state.

The fact that the rank-2 GHZ state ($\alpha=90^{o}$) has the highest resistance to white noise, needs a closer look. Note that the state is effectively a three-{\em qubit} one. If one uses observables T, which have eigenvectors which are unbiased with respect to the local computational bases, all that disappears, such states have classical correlations. The high white noise resistance  stems from the fact that part of the noise affects only a ``blind" third detector, which is in the case of zero noise not registering any counts, because the observables, for which we get this low noise value, are such that their non trivial  eigenstates are superpositions of $|1\rangle$ and $|2\rangle$, while the third one is always $|0\rangle$ (we have effectively a two-qubit coincidence interference). That is, the effect seems to be due to the fact that white noise is a good model of ``dark counts", rather than noise due to decoherence or distortion in the channels leading to the detectors.

Note also the peculiar behavior of resistance to product noise of generalized  GHZ states, which are very close to the ``three-qubit" one, Fig. \ref{Fig:prod} and the caption. If  $\alpha<90^{o}$, but very close to $90^{o}$ (up to our numerical accuracy), we have an almost perfect resistance to the product noise. Suddenly, for $\alpha=90^{o}$ one has the critical value of $0.5000$ (almost as in the case of the three-qutrit GHZ state $\ket{\psi_{sym}}$, $v_{crit}=0.5263$). This means that  different tight Bell inequalities are optimal for the  ``three-qubit" GHZ state and for  the irreducibly three-qutrit states very close to it. This phenomenon, as the product noise could in a way model a distortion in channels, might be useful in transmitting  the GHZ correlations.

\ack
This work forms part of the Foundation for Polish Science TEAM project co-financed by the EU European Regional Development Fund.
J.R. is supported by a NCBiR-CHIST-ERA Project QUASAR. 
WL and M\.Z are  supported by Polish Ministry of Science and Higher Education Grant no. IdP2011 000361.

\section*{Appendix: Measurement parameters corresponding to $v({\rm parameters}) = v_{crit}$}

\twocolumn

\begin{enumerate}

\item Parametrization of the measurements 

\begin{itemize}
\item  {\em U(3) measurements} --  the full family of observables, defined
by U(3)-type unitary transformations. For the $m$th observer and the $n$th measurement setting the transformation is given by (see e.g \cite{TILMA}):
\begin{eqnarray}
\hspace{-2.5cm}&&U(3)^{m,n} = e^{i \lambda_3\alpha_0^{m,n}} e^{i\lambda_2\alpha_1^{m,n}} e^{i\lambda_3\alpha_2^{m,n}} e^{i\lambda_5\alpha_3^{m,n}} \nonumber \\ 
\hspace{-2.5cm}&&\times e^{i\lambda_3\alpha_4^{m,n}} e^{i\lambda_2\alpha_5^{m,n}} e^{i\lambda_3\alpha_6^{m,n}} e^{i\lambda_8\alpha_7^{m,n}}, \nonumber 
\end{eqnarray}
where $\lambda_i$ denote the Gell-Mann matrices.

\item  {\em T  mesurements} -- a subset of the above observables which are defined by unitary
transformations, starting form the computational basis, which use one unbiased three-
port beam splitter performing a unitary transformation given by the matrix 
$U_{kl} = e^{2 \pi i k l/3}$, and a set of phase shifters (for the $m$th observer and the $n$th measurement setting) $(\Phi^{m,n})_{kl} = \delta_{k,l} \phi_k^{m,n}$ in front of it ($k=0,1,2$). 
\end{itemize}

\item Optimal measurement parameters

\begin{enumerate}

\item State: $|\psi_{sym}\rangle$
\begin{itemize}
\item Type of mesurements: T
\item No. of measur. settings: 2
\item $v_{crit} =0.6$ 
\item parameters: \\
$\phi_0^{m,0}=3.1357580835051193; \\ \phi_0^{m,1}=4.9708784423681580;\\ \phi_1^{m,0}=1.0414443119091470;\\ \phi_1^{m,1}=4.9721261843082081;\\ \phi_2^{m,0}=5.5757854568768224;\\ \phi_2^{m,1}=3.2243818908364767\\ (m=0,1,2);$
\end{itemize}

\item State: $|\psi_{sym}\rangle$
\begin{itemize}
\item Type of mesurements: T
\item No. of measur. settings: 3
\item $v_{crit} =0.5$ 
\item parameters: \\
$\phi_0^{m,0}=2.0132478996893455;\\
\phi_0^{m,1}=5.4472235105464843;\\
\phi_0^{m,2}=6.0968810475923743;\\
\phi_1^{m,0}=5.4978525717510918;\\
\phi_1^{m,1}=5.8001994115278581;\\
\phi_1^{m,2}=1.2128355321845690;\\
\phi_2^{m,0}=0.9708052150061472;\\
\phi_2^{m,1}=4.4016389085254533;\\
\phi_2^{m,2}=2.9469488719772410;\\
(m=0,1,2);$
\end{itemize}

\item State: $|\psi_{sym}\rangle$
\begin{itemize}
\item Type of mesurements: U(3)
\item No. of measur. settings: 2
\item $v_{crit} =0.5263$ 
\item parameters: \\
$
\alpha_0^{m,0}=0.1095979254970793;\\
\alpha_0^{m,1}=3.3097248621683324;\\
\alpha_1^{m,0}=1.6449871624952401;\\
\alpha_1^{m,1}=5.2344494341225953;\\
\alpha_2^{m,0}=5.6347997620619674;\\
\alpha_2^{m,1}=0.7836304069095931;\\
\alpha_3^{m,0}=5.5874048717425708;\\
\alpha_3^{m,1}=4.7515674960770875;\\
\alpha_4^{m,0}=3.1407663841155311;\\
\alpha_4^{m,1}=1.5704416776930992;\\
\alpha_5^{m,0}=4.7870141478002921;\\
\alpha_5^{m,1}=1.2268620656547207;\\
\alpha_6^{m,0}=1.2135652845463178;\\
\alpha_6^{m,1}=3.1423279172669081;\\
\alpha_7^{m,0}=1.0386933887412846;\\
\alpha_7^{m,1}=2.2265752989927949;\\ (m=0,1,2);$
\end{itemize}

\item State: $|\psi_{asym}\rangle$
\begin{itemize}
\item Type of mesurements: T
\item No. of measur. settings: 2
\item $v_{crit} =0.59308$ 
\item parameters: \\
$\phi_0^{0,0}=4.8704051315456711;\\
\phi_0^{0,1}=0.3552175123860086;\\
\phi_0^{1,0}=1.0093537634178003;\\
\phi_0^{1,1}=2.6170872946665042;\\
\phi_0^{2,0}=1.7489643378140376;\\
\phi_0^{2,1}=5.1780638544894897;\\
\phi_1^{0,0}=0.3422452951514470;\\
\phi_1^{0,1}=4.2067267401700938;\\
\phi_1^{1,0}=4.5396731790544775;\\
\phi_1^{1,1}=6.1530106453243141;\\
\phi_1^{2,0}=3.7971198031734490;\\
\phi_1^{2,1}=3.0310064080448540;\\
\phi_2^{0,0}=5.4509559415874209;\\
\phi_2^{0,1}=3.0237821542589116;\\
\phi_2^{1,0}=3.5707386359149158;\\
\phi_2^{1,1}=3.0818123177307188;\\
\phi_2^{2,0}=5.9387726991200154;\\
\phi_2^{2,1}=5.1829394347970075;$
\end{itemize}

\item State: $|\psi_{asym}\rangle$
\begin{itemize}
\item Type of mesurements: U(3)
\item No. of measur. settings: 2
\item $v_{crit} =0.5192$ 
\item parameters: \\
$\alpha_0^{0,0}=0.3815699296323373;\\
\alpha_0^{0,1}=1.5480398197862630;\\
\alpha_0^{1,0}=2.0910510773242197;\\
\alpha_0^{1,1}=0.4953180043520519;\\
\alpha_0^{2,0}=4.9562014295553798;\\
\alpha_0^{2,1}=3.1499839702264225;\\
\alpha_1^{0,0}=5.5634530591517386;\\
\alpha_1^{0,1}=5.5834933406885785;\\
\alpha_1^{1,0}=3.2562859499511219;\\
\alpha_1^{1,1}=2.5388671006701009;\\
\alpha_1^{2,0}=2.1622537062610681;\\
\alpha_1^{2,1}=3.6505423283612335;\\
\alpha_2^{0,0}=3.2391908173208228;\\
\alpha_2^{0,1}=0.7869708606856278;\\
\alpha_2^{1,0}=0.7851908628306062;\\
\alpha_2^{1,1}=2.7169250264581111;\\
\alpha_2^{2,0}=5.4822036426809220;\\
\alpha_2^{2,1}=5.9445930142799188;\\
\alpha_3^{0,0}=0.4692043527096480;\\
\alpha_3^{0,1}=2.6764600123751059;\\
\alpha_3^{1,0}=4.9428088836238038;\\
\alpha_3^{1,1}=2.0470710881373217;\\
\alpha_3^{2,0}=0.4773163765524995;\\
\alpha_3^{2,1}=0.2370881315684034;\\
\alpha_4^{0,0}=0.0029648535983000;\\
\alpha_4^{0,1}=4.7129099980672482;\\
\alpha_4^{1,0}=2.0567728842770139;\\
\alpha_4^{1,1}=3.1397516471214728;\\
\alpha_4^{2,0}=0.0073949671777779;\\
\alpha_4^{2,1}=3.1398347033677245;\\
\alpha_5^{0,0}=3.4761808423356051;\\
\alpha_5^{0,1}=0.5039095751421377;\\
\alpha_5^{1,0}=5.2159458749139924;\\
\alpha_5^{1,1}=5.6978725838787296;\\
\alpha_5^{2,0}=1.1728454158269070;\\
\alpha_5^{2,1}=3.7521077606199089;\\
\alpha_6^{0,0}=2.3605096818119420;\\
\alpha_6^{0,1}=6.2779388174414947;\\
\alpha_6^{1,0}=3.1526653412808918;\\
\alpha_6^{1,1}=5.9176717084434323;\\
\alpha_6^{2,0}=3.1245589449749205;\\
\alpha_6^{2,1}=5.5510950602422966;\\
\alpha_7^{0,0}=2.6191197925387040;\\
\alpha_7^{0,1}=6.2483921100411912;\\
\alpha_7^{1,0}=6.1017265255235866;\\
\alpha_7^{1,1}=3.5169857773730828;\\
\alpha_7^{2,0}=2.7167060771422324;\\
\alpha_7^{2,1}=4.3439539761778221;$
\end{itemize}

\item State: $|\psi(90^{o})\rangle$
\begin{itemize}
\item Type of mesurements: U(3)
\item No. of measur. settings: 2
\item $v_{crit} =0.4256$ 
\item parameters: \\
$
\alpha_0^{m,0}=3.6886338613072405;\\
\alpha_0^{m,1}=5.8366938512619839;\\
\alpha_1^{m,0}=1.6211198704953957;\\
\alpha_1^{m,1}=4.7625316234077975;\\
\alpha_2^{m,0}=4.7095912437869316;\\
\alpha_2^{m,1}=3.1453019521360912;\\
\alpha_3^{m,0}=2.1457023024875035;\\
\alpha_3^{m,1}=4.0023194994893032;\\
\alpha_4^{m,0}=5.5011513790373456;\\
\alpha_4^{m,1}=1.5662338757924097;\\
\alpha_5^{m,0}=2.0192381176418808;\\
\alpha_5^{m,1}=5.5667973575541190;\\
\alpha_6^{m,0}=4.7089657665238907;\\
\alpha_6^{m,1}=0.7830582351695802;\\
\alpha_7^{m,0}=3.1791741530555573;\\
\alpha_7^{m,1}=0.2311220570800439;\\ (m=0,1,2)$
\end{itemize}

\item State: $|D_3^1\rangle$
\begin{itemize}
\item Type of mesurements: U(3)
\item No. of measur. settings: 2
\item $v_{crit} =0.4993$ 
\item parameters: \\
$\alpha_0^{m,0}=5.8305236734139969;\\
\alpha_0^{m,1}=1.8572002169969064;\\
\alpha_1^{m,0}=4.8576348242255802;\\
\alpha_1^{m,1}=3.8401272975333143;\\
\alpha_2^{m,0}=3.7344276859596408;\\
\alpha_2^{m,1}=3.8583650549459616;\\
\alpha_3^{m,0}=3.1519307422507974;\\
\alpha_3^{m,1}=0.7973163086361438;\\
\alpha_4^{m,0}=6.2816527253152614;\\
\alpha_4^{m,1}=0.0033809205758466;\\
\alpha_5^{m,0}=6.1701588001391441;\\
\alpha_5^{m,1}=4.9438678203500359;\\
\alpha_6^{m,0}=2.4432010868897400;\\
\alpha_6^{m,1}=5.6698959751796139;\\
\alpha_7^{m,0}=3.7303639159087960;\\
\alpha_7^{m,1}=5.1105982584756164; \\(m=0,1,2)$\\
\end{itemize}

\item State: $|D_3^1\rangle$
\begin{itemize}
\item Type of mesurements: U(3)
\item No. of measur. settings: 3
\item $v_{crit} =0.4866$ 
\item parameters: \\
$\alpha_0^{m,0}=2.7232603603418486;\\
\alpha_0^{m,1}=5.6540019179495342;\\
\alpha_0^{m,2}=3.0406378641532519;\\
\alpha_1^{m,0}=4.6966544548263212;\\
\alpha_1^{m,1}=2.2713520583753972;\\
\alpha_1^{m,2}=4.5202112991131065;\\
\alpha_2^{m,0}=4.0045554557022021;\\
\alpha_2^{m,1}=6.2755952887395408;\\
\alpha_2^{m,2}=4.2789122867291658;\\
\alpha_3^{m,0}=2.0314414618819603;\\
\alpha_3^{m,1}=0.0629107166988518;\\
\alpha_3^{m,2}=1.4049139354953630;\\
\alpha_4^{m,0}=3.1469202614276188;\\
\alpha_4^{m,1}=1.5692666086084461;\\
\alpha_4^{m,2}=3.1428136191246603;\\
\alpha_5^{m,0}=1.1233583013497295;\\
\alpha_5^{m,1}=4.8571745614205746;\\
\alpha_5^{m,2}=4.3053423659738401;\\
\alpha_6^{m,0}=3.4978395105343481;\\
\alpha_6^{m,1}=1.5562479293525660;\\
\alpha_6^{m,2}=0.1400728400743489;\\
\alpha_7^{m,0}=4.4474435696017149;\\
\alpha_7^{m,1}=2.0194397074192847;\\
\alpha_7^{m,2}=2.3049445813197740;\\ (m=0,1,2)$
\end{itemize}

\item State: $|D_3^2\rangle$
\begin{itemize}
\item Type of mesurements: U(3)
\item No. of measur. settings: 2
\item $v_{crit} =0.5368$ 
\item parameters: \\
$\alpha_0^{m,0}=1.8758643191411610;\\
\alpha_0^{m,1}=4.1409050341582008;\\
\alpha_1^{m,0}=0.7924995327971064;\\
\alpha_1^{m,1}=3.8730567287633670;\\
\alpha_2^{m,0}=1.8330371503673553;\\
\alpha_2^{m,1}=4.4200032172042159;\\
\alpha_3^{m,0}=2.2636790381492244;\\
\alpha_3^{m,1}=5.2447011775792660;\\
\alpha_4^{m,0}=2.5388781667799236;\\
\alpha_4^{m,1}=1.4523514914469635;\\
\alpha_5^{m,0}=2.4846001602088523;\\
\alpha_5^{m,1}=5.7463176090768489;\\
\alpha_6^{m,0}=1.3229663962829663;\\
\alpha_6^{m,1}=2.9761316547184453;\\
\alpha_7^{m,0}=0.5255778354059855;\\
\alpha_7^{m,1}=4.0984854031488211; \\(m=0,1,2)
$
\end{itemize}

\item State: $|D_3^2\rangle$
\begin{itemize}
\item Type of mesurements: U(3)
\item No. of measur. settings: 3
\item $v_{crit} =0.5285$ 
\item parameters: \\
$\alpha_0^{m,0}=1.3670735040071715;\\
\alpha_0^{m,1}=2.3087560666611431;\\
\alpha_0^{m,2}=1.9763919904705367;\\
\alpha_1^{m,0}=5.6657195902047039;\\
\alpha_1^{m,1}=5.2103437584964949;\\
\alpha_1^{m,2}=0.9810630121506206;\\
\alpha_2^{m,0}=1.6732431588777543;\\
\alpha_2^{m,1}=1.0435974227940708;\\
\alpha_2^{m,2}=5.9824237497917316;\\
\alpha_3^{m,0}=1.3731716237604459;\\
\alpha_3^{m,1}=0.9344779956564189;\\
\alpha_3^{m,2}=6.1552276757390114;\\
\alpha_4^{m,0}=1.3768336845235378;\\
\alpha_4^{m,1}=2.4460705340209690;\\
\alpha_4^{m,2}=4.9565132212802769;\\
\alpha_5^{m,0}=1.1184418576593886;\\
\alpha_5^{m,1}=2.2560931968089299;\\
\alpha_5^{m,2}=3.2935221736938378;\\
\alpha_6^{m,0}=3.8767810592199692;\\
\alpha_6^{m,1}=5.8877149141539231;\\
\alpha_6^{m,2}=4.3853875102352227;\\
\alpha_7^{m,0}=1.5974205190343056;\\
\alpha_7^{m,1}=1.4511777939272612;\\
\alpha_7^{m,2}=4.2474134108578419; \\ (m=0,1,2)$
\end{itemize}

\item State: $|D_3^3\rangle$
\begin{itemize}
\item Type of mesurements: U(3)
\item No. of measur. settings: 2
\item $v_{crit} =0.5338$ 
\item parameters: \\
$\alpha_0^{m,0}=2.8224970414401969;\\
\alpha_0^{m,1}=1.3945081719827259;\\
\alpha_1^{m,0}=0.3620480228225115;\\
\alpha_1^{m,1}=2.5664813099128025;\\
\alpha_2^{m,0}=3.1503587455366469;\\
\alpha_2^{m,1}=3.1420132460507144;\\
\alpha_3^{m,0}=4.6013997034433434;\\
\alpha_3^{m,1}=5.5399211163673021;\\
\alpha_4^{m,0}=3.9284858863642333;\\
\alpha_4^{m,1}=3.9266639246641954;\\
\alpha_5^{m,0}=1.7713962812616508;\\
\alpha_5^{m,1}=4.1286259824114246;\\
\alpha_6^{m,0}=2.3644210167477251;\\
\alpha_6^{m,1}=5.4983819930868911;\\
\alpha_7^{m,0}=6.2432939558270792;\\
\alpha_7^{m,1}=3.3165492879895240; \\(m=0,1,2);
$
\end{itemize}

\item State: $|D_3^3\rangle$
\begin{itemize}
\item Type of mesurements: U(3)
\item No. of measur. settings: 3
\item $v_{crit} =0.5108$ 
\item parameters: \\
$\alpha_0^{m,0}=2.5260062350462800;\\
\alpha_0^{m,1}=4.2772116833005605;\\
\alpha_0^{m,2}=1.7697583196949682;\\
\alpha_1^{m,0}=4.4806168431512639;\\
\alpha_1^{m,1}=5.8668104741271208;\\
\alpha_1^{m,2}=1.6116761381164195;\\
\alpha_2^{m,0}=0.0036830136289101;\\
\alpha_2^{m,1}=0.0025793058306753;\\
\alpha_2^{m,2}=0.0021331983540849;\\
\alpha_3^{m,0}=1.0544832107238367;\\
\alpha_3^{m,1}=5.0042519161480383;\\
\alpha_3^{m,2}=0.7532655373841105;\\
\alpha_4^{m,0}=0.7917633790698999;\\
\alpha_4^{m,1}=1.0796196634605915;\\
\alpha_4^{m,2}=3.9290866194808052;\\
\alpha_5^{m,0}=3.7024805723680396;\\
\alpha_5^{m,1}=4.4456565227896929;\\
\alpha_5^{m,2}=1.3689234786715838;\\
\alpha_6^{m,0}=0.7887334735135703;\\
\alpha_6^{m,1}=4.7103568217090839;\\
\alpha_6^{m,2}=0.7845024411421854;\\
\alpha_7^{m,0}=4.0028466899417898;\\
\alpha_7^{m,1}=1.2358495996142038;\\
\alpha_7^{m,2}=5.2481044645418180; \\(m=0,1,2)
$
\end{itemize}

\end{enumerate}

\end{enumerate}

\onecolumn

\end{document}